\documentclass[%
superscriptaddress,
twocolumn,
prb,
floatfix,
]{revtex4-1}

\usepackage{graphicx}
\usepackage{bm}
\usepackage{amsmath}
\usepackage{hyperref} \hypersetup{colorlinks=true,citecolor=blue,linkcolor=blue,urlcolor=blue}
\usepackage{bbm}
\usepackage[export]{adjustbox}
\usepackage[dvipsnames]{xcolor}
\usepackage{multirow}
\usepackage{array}      

\DeclareMathOperator*{\argmin}{arg\,min}
\DeclareMathOperator*{\avg}{avg}
\DeclareMathOperator*{\std}{std}

\bibpunct{[}{]}{,}{n}{}{}   

\def\AGL{{\small AGL}}
\def\ML{{\small ML}}
\def\AEL{{\small AEL}}
\def\AFLOW{{\small AFLOW}}
\def\ICSD{{\small ICSD}}

\def\QMSPR{{\small QMSPR}}
\def\QSAR{{\small QSAR}}
\def\GBDT{{\small GBDT}}
\def\PLMF{{\small PLMF}}
\def\GW{{\small GW}}
\def\DFT{{\small DFT}}
\def\RMSE{{\small RMSE}}
\def\MAE{{\small MAE}}
\def\AUC{{\small AUC}}
\def\ROC{{\small ROC}}
\def\CCR{{\small CCR}}
\def\MOF{{\small MOF}}

\def\sDebye{{\substack{\scalebox{0.6}{D}}}}
\def\sD{\sDebye}

\def\sVRH{{\substack{\scalebox{0.6}{VRH}}}}

\def\sBG{{\substack{\scalebox{0.6}{BG}}}}

\def\sP{{\substack{\scalebox{0.6}{P}}}}
\def\sp{{\substack{\scalebox{0.6}{p}}}}
\def\sV{{\substack{\scalebox{0.6}{V}}}}
\def\sB{{\substack{\scalebox{0.6}{B}}}}
\def\sa{{\substack{\scalebox{0.6}{a}}}}
\def\sbond{{\substack{\scalebox{0.6}{bond}}}}
\def\svapor{{\substack{\scalebox{0.6}{vapor}}}}
\def\satom{{\substack{\scalebox{0.6}{atom}}}}
\def\sfusion{{\substack{\scalebox{0.6}{fusion}}}}
\def\smolar{{\substack{\scalebox{0.6}{molar}}}}
\def\scov{{\substack{\scalebox{0.6}{cov}}}}
\def\seff{{\substack{\scalebox{0.6}{eff}}}}

\newcommand\electronicTotal{26,674}
\newcommand\insulatorTotal{13,812}
\newcommand\metalTotal{12,862}
\newcommand\thermoTrainingTotal{2,829}
\newcommand\thermoTestTotal{770}

\begin{document}

\title{
  Universal Fragment Descriptors for Predicting Properties of Inorganic Crystals
}

\author{Olexandr Isayev}
\email{olexandr@olexandrisayev.com}
\affiliation{Laboratory for Molecular Modeling, Division of Chemical Biology and Medicinal Chemistry, 
  UNC Eshelman School of Pharmacy, University of North Carolina, Chapel Hill, NC, 27599, USA}
\author{Corey Oses}%
\affiliation{Center for Materials Genomics, Duke University, Durham, NC 27708, USA}%
\author{Cormac Toher}%
\affiliation{Center for Materials Genomics, Duke University, Durham, NC 27708, USA}%
\author{Eric Gossett}%
\affiliation{Center for Materials Genomics, Duke University, Durham, NC 27708, USA}%
\author{Stefano Curtarolo}
\email{stefano@duke.edu}
\affiliation{Center for Materials Genomics, Duke University, Durham, NC 27708, USA}%
\affiliation{Materials Science, Electrical Engineering, Physics and Chemistry, Duke University, Durham, NC 27708, USA}%
\author{Alexander Tropsha}
\affiliation{Laboratory for Molecular Modeling, Division of Chemical Biology and Medicinal Chemistry, 
  UNC Eshelman School of Pharmacy, University of North Carolina, Chapel Hill, NC, 27599, USA}

\date{\today}

\begin{abstract}
  Historically, materials discovery has been
  driven by a laborious trial-and-error process.
  The growth of materials databases and emerging informatics approaches finally offer the opportunity 
  to transform this practice into data- and knowledge-driven rational design.
  By using data from the \AFLOW\ repository for high-throughput \textit{ab-initio} calculations, we have generated
  \underline{Q}uantitative \underline{M}aterials \underline{S}tructure-\underline{P}roperty \underline{R}elationship (\QMSPR) 
  models to predict eight critical electronic and thermomechanical materials properties, 
  such as the metal/insulator classification, band gap energy, bulk
  and shear moduli, Debye temperature, and heat capacity.
  The prediction accuracy obtained with these \QMSPR\ models approaches training data 
  for virtually any stoichiometric inorganic crystalline material.
  The success and universality of these models is attributed to the construction of new 
  materials descriptors---referred to as the universal \underline{P}roperty-\underline{L}abeled \underline{M}aterials 
  \underline{F}ragments (\PLMF).  
  The representation requires only minimal structural input and affords straightforward 
  model interpretation in terms of simple heuristic design rules that guide rational materials design.  
  This study  demonstrates the power of materials informatics to dramatically accelerate the search for new materials.
\end{abstract}

\maketitle


\section{Introduction}
Advances in materials science are often slow and fortuitous~\cite{nmatHT}.
Coupling the field's combinatorial challenges with the demanding efforts required for materials characterization
makes progress uniquely difficult.
The number of materials currently characterized, either experimentally~\cite{ICSD,ICSD3} or 
computationally~\cite{curtarolo:art75,APL_Mater_Jain2013,Saal_JOM_2013,Hachmann_JPCL_2011}, 
pales in comparison to the anticipated potential diversity.
Only considering naturally occurring elements, 9,000 crystal structure prototypes~\cite{ICSD,ICSD3}, 
and stoichiometric compositions, there are roughly $3 \times 10^{11}$ potential quaternary compounds and
$10^{13}$ quinary combinations.
Indeed, it has been estimated that the total number of theoretical materials can be as large as 
$10^{100}$~\cite{Walsh_NChem_2015}.
Exacerbating the issue, standard materials characterization practices, such as calculating the band structure,  
can become quite expensive when considering 
finite-size scaling, charge corrections~\cite{Castleton_PRB_2006},
and going beyond standard density functional theory (\DFT) with Green's function methods such as the fully 
self-consistent \GW\ approximation~\cite{Lindgren_BetheSalpeter_2011,vanSchilfgaarde_PRL_2006,Stan_JCP_2009}.
Ultimately, brute force exploration of this search space, 
even in high-throughput fashion~\cite{nmatHT,koinuma_nmat_review2004,Potyrailo2005}, is entirely impractical.

To circumvent the issue, many knowledge-based structure-property 
relationships have been conjectured over the years to aid in the search for novel functional materials---ranging
from the simplest empirical relationships~\cite{Mizutani_HR_2011} 
to complex advanced 
models~\cite{curtarolo:art94,Ghiringhelli_PRL_2015,PyzerKnapp_AdFM_2015,Rajan_ARMR_2015,curtarolo:art85,curtarolo:art84,curtarolo:art120,Furmanchuk_RSCA_2016}.
For instance, many (semi-)empirical rules have been developed to predict 
band gap energies, such as those incorporating 
(optical~\cite{Duffy_JCP_1977}) electronegativity~\cite{DiQuarto_JPCB_1997}.
More sophisticated \underline{M}achine \underline{L}earning (\ML) models were also developed for chalcopyrite 
semiconductors~\cite{Zeng_ChemMat_2002,rajan:ref3}, perovskites~\cite{Pilania_SR_2016}, and binary 
compounds~\cite{Gu_SSS_2006}.
Unfortunately, many of these models are limited to a single family of materials, 
with narrow applicability outside of their training scope.

The development of such structure-property relationships has become an integral practice in the drug industry,
which faces a similar combinatorial challenge.
The number of potential organic molecules is estimated to be anywhere between $10^{13}$ to $10^{180}$~\cite{Gorse_CTMC_2006}.
In computational medicinal chemistry, \underline{Q}uantitative \underline{S}tructure-\underline{A}ctivity 
\underline{R}elationship (\QSAR) modeling coupled with 
virtual screening of chemical libraries have been largely successfully in the discovery of 
novel bioactive compounds~\cite{Baskin_VirScreen_2008,Kitchen_NRDD_2004}.
This parallel with drug innovation suggests an opportunity to develop and employ similar modeling
approaches to materials discovery.

Here, we introduce novel fragment descriptors of materials structure.
The combination of these descriptors with \ML\ approaches affords the development of universal models 
capable of accurate prediction of properties for virtually any stoichiometric inorganic crystalline material.
First, the algorithm for descriptor generation is described, along with implementation of
\ML\ methods for \underline{Q}uantitative \underline{M}aterials \underline{S}tructure-\underline{P}roperty 
\underline{R}elationship (\QMSPR) modeling.
Next, the effectiveness of this approach is assessed through prediction of eight critical electronic
and thermomechanical properties of materials,
including the metal/insulator classification, band gap energy, bulk and shear moduli, 
Debye temperature, heat capacities (at constant pressure and volume), and thermal expansion coefficient.
The impact and interaction among the most significant descriptors as determined
by the \ML\ algorithms are highlighted.
As a proof-of-concept, the \QMSPR\ models are then employed to predict thermomechanical properties
for compounds previously uncharacterized, and the predictions are validated via 
the \AEL-\AGL\ integrated framework~\cite{curtarolo:art96, Toher_AEL_2016}.
Such predictions are of particular value as proper calculation pathways for thermomechanical properties 
in the most efficient scenarios still require analysis of multiple \DFT-runs, elevating the cost
of already expensive calculations.
Finally, \ML-predictions and calculations are both compared to experimental values which ultimately corroborate the validity of the approach.

Other investigations have predicted a subset of the target properties
discussed here
by building \ML\ approaches where computationally obtained quantities,
such as the cohesive energy, formation energy, and energy above the
convex hull,
are the part of the input data~\cite{deJong_SR_2016}.
The approach presented here is orthogonal.
Once trained, our proposed models achieve comparable accuracies without the need of
further \textit{ab-initio} data.
All necessary input properties are either tabulated or derived directly
from the geometrical structures.
There are advantages:
\textit{(i)} \textit{a priori}, after the training, no further calculations need
to be performed,
\textit{(ii)} \textit{a posteriori}, the modeling framework becomes
independent of the source or nature of the training data,
\textit{e.g.}, calculated \textit{vs.} experimental.
The latter allows for rapid extension of predictions to online
applications---given the geometry of a cell and the species involved, eight \ML\
predicted properties are returned
(\href{http://aflow.org/aflow-ml}{aflow.org/aflow-ml}).

\begin{figure*}[t!]
  \includegraphics[width=\textwidth]{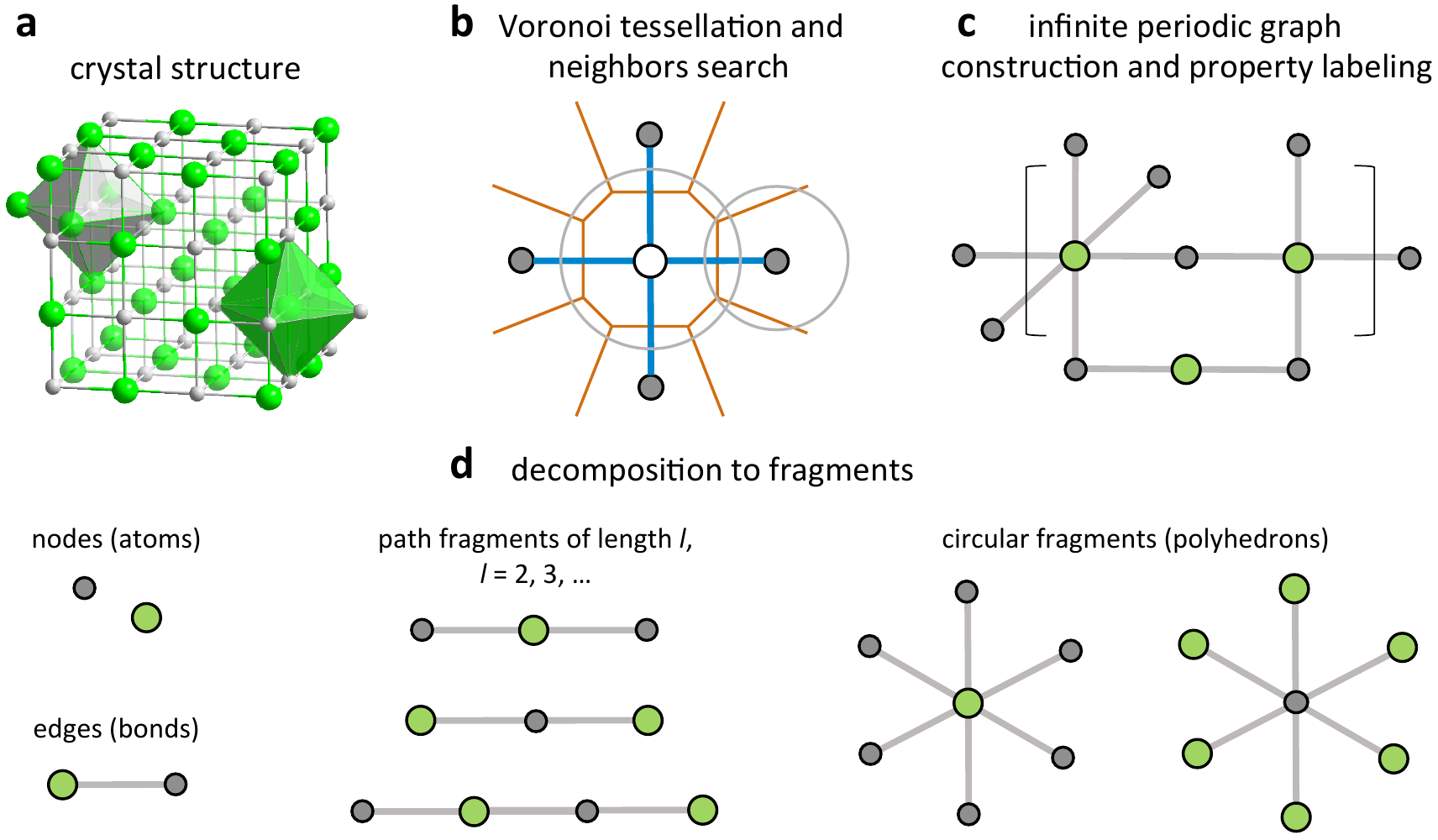}
  \caption{\small 
    \textbf{Schematic representing the construction of the \underline{P}roperty-\underline{L}abeled 
    \underline{M}aterials \underline{F}ragments (\PLMF).}
    The crystal structure (\textbf{a}) is analyzed for atomic neighbors (\textbf{b}) via Voronoi tessellation.
    After property labeling, the resulting periodic graph (\textbf{c}) is decomposed into simple
    subgraphs (\textbf{d}).}
  \label{figure1}
\end{figure*}

\section{Methods}
\label{sec:methods}

\noindent \textbf{Data preparation.}
Two independent datasets were prepared for the creation and validation of the \ML\ models.
The \textit{training set} includes 
electronic~\cite{curtarolo:art75,curtarolo:art65,curtarolo:art58,curtarolo:art67,curtarolo:art87,curtarolo:art49} 
and thermomechanical properties~\cite{curtarolo:art96, Toher_AEL_2016} for a broad diversity of
compounds already characterized in the \AFLOW\ database.
This set is used to build and analyze the \ML\ models, one model per property. 
The constructed thermomechanical models are then employed to make predictions of previously uncharacterized compounds in the \AFLOW\ database.
Based on these predictions and consideration of computational cost, several compounds are selected to validate the models' predictive
power.
These compounds and their newly computed properties define the \textit{test set}.
The compounds used in both datasets are specified in the Supplementary Information.

\noindent\textbf{Training set.}
Band gap energy data for 49,934 materials were extracted from the \AFLOW\ 
repository~\cite{curtarolo:art75,curtarolo:art65,curtarolo:art58,curtarolo:art67,curtarolo:art87,curtarolo:art49}, representing approximately 
60\% of the known stoichiometric inorganic crystalline materials listed in the 
\underline{I}norganic \underline{C}rystal \underline{S}tructure \underline{D}atabase (\ICSD)~\cite{ICSD,ICSD3}.  
While these band gap energies are generally underestimated with respect to experimental 
values~\cite{Perdew_IJQC_1985}, \DFT+$U$ is robust enough to 
differentiate between metallic (no $E_\sBG$) and insulating $\left(E_\sBG\!>\!0\right)$ systems~\cite{curtarolo:art104}.
Additionally, errors in band gap energy prediction are typically systematic.  
Therefore, the band gap energy values can be corrected \textit{ad-hoc} with fitting 
schemes~\cite{Yazyev_PRB_2012,Zheng_PRL_2011}.
Prior to model development, both \ICSD\ and \AFLOW\ data were curated:
duplicate entries, erroneous structures, and ill-converged calculations were corrected or removed.  
Noble gases crystals are not considered.
The final dataset consists of \electronicTotal\ unique materials (\metalTotal\ with no $E_\sBG$
and \insulatorTotal\ with $E_\sBG\!>\!0$), 
covering the seven lattice systems, 230 space groups, and 83 elements 
(H-Pu, excluding noble gases, Fr, Ra, Np, At, and Po).
All referenced \DFT\ calculations were performed with the \underline{G}eneralized \underline{G}radient \underline{A}pproximation 
(GGA) PBE~\cite{PBE}
exchange-correlation functional and \underline{p}rojector-\underline{a}ugmented \underline{w}avefunction (PAW) 
potentials~\cite{PAW,kresse_vasp_paw} according to the
\AFLOW\ Standard for \underline{H}igh-\underline{T}hroughput (HT) Computing~\cite{curtarolo:art104}.
The Standard ensures reproducibility of the data, and provides visibility/reasoning for any parameters 
set in the calculation, such as accuracy thresholds, calculation
pathways, and mesh dimensions. 

Thermomechanical properties data for just over 3,000 materials were extracted from the \AFLOW\ 
repository~\cite{Toher_AEL_2016}.
These properties include the bulk modulus, shear modulus, Debye temperature, heat capacity at constant pressure,
heat capacity at constant volume, and thermal expansion coefficient, and were
calculated using the \AEL-\AGL\ integrated framework~\cite{curtarolo:art96, Toher_AEL_2016}.
The \AEL\ ({\small \underline{A}FLOW} \underline{E}lasticity \underline{L}ibrary) 
method~\cite{Toher_AEL_2016} applies a set of independent normal and shear strains to the structure, and then fits the calculated stress 
tensors to obtain the elastic constants~\cite{curtarolo:art100}. 
These can then be used to calculate the elastic moduli in 
the Voigt and Reuss approximations, as well as the Voigt-Reuss-Hill (VRH) averages which are the values of the bulk and 
shear moduli modeled in this work. 
The \AGL\ ({\small \underline{A}FLOW} {\small \underline{G}IBBS} \underline{L}ibrary) method~\cite{curtarolo:art96} 
fits the energies from a set of isotropically 
compressed and expanded volumes of a structure to a quasiharmonic Debye-Gr{\"u}neisen model~\cite{Blanco_CPC_GIBBS_2004} 
to obtain thermomechanical 
properties, including the bulk modulus, Debye temperature, heat capacity, and thermal expansion coefficient. 
\AGL\ has been 
combined with \AEL\ in a single workflow, so that it can utilize the Poisson ratios obtained from \AEL\ to improve the 
accuracy of the thermal properties predictions~\cite{Toher_AEL_2016}.
After a similar curation of ill-converged calculations, the final dataset consists of 
\thermoTrainingTotal\ materials.
It covers the seven lattice systems, includes unary, binary, and ternary compounds, and 
spans broad ranges of each thermomechanical property, including
high thermal conductivity systems such as C (\ICSD\ \#182729), BN (\ICSD\ \#162874), BC$_{5}$ (\ICSD\ \#166554), 
CN$_{2}$ (\ICSD\ \#247678), MnB$_{2}$ (\ICSD\ \#187733), and SiC (\ICSD\ \#164973), as well as 
low thermal conductivity systems such as Hg$_{33}$(Rb,K)$_{3}$ (\ICSD\ \#410567 and \#410566), 
Cs$_{6}$Hg$_{40}$ (\ICSD\ \#240038), Ca$_{16}$Hg$_{36}$ (\ICSD\ \#107690), CrTe (\ICSD\ \#181056), 
and Cs (\ICSD\ \#426937).
{Many of these systems additionally exhibit extreme values of the bulk and shear moduli,
such as C (high bulk and shear moduli) and Cs (low bulk and shear moduli).
Interesting systems such as
RuC (\ICSD\ \#183169) and NbC (\ICSD\ \#189090)
with a high bulk modulus ($B_\sVRH$ = 317.92 GPa, 263.75 GPa) but 
low shear modulus ($G_\sVRH$ = 16.11 GPa, 31.86 GPa)
also populate the set.
}

\noindent\textbf{Test set.}
While nearly all \ICSD\ compounds are characterized electronically within the \AFLOW\ database,
most have not been characterized thermomechanically due to the added computational cost.
This presented an opportunity to validate the \ML\ models.
Of the remaining compounds, several were prioritized for immediate characterization via
the \AEL-\AGL\ integrated framework~\cite{curtarolo:art96, Toher_AEL_2016}.
In particular, focus was placed on systems predicted to have a large bulk modulus, as this property
is expected to scale well with the other aforementioned thermomechanical 
properties~\cite{curtarolo:art96, Toher_AEL_2016}.
The set also includes various other small cell, high symmetry systems expected to span the full 
applicability domains of the models.
This effort resulted in the characterization of \thermoTestTotal\ additional compounds.

\noindent\textbf{Universal Property-Labeled Materials Fragments.}
Many cheminformatics investigations have demonstrated the critical importance of 
molecular descriptors, which are known to 
influence model accuracy more than the choice of the 
\ML\ algorithm~\cite{Young_MI_2012,Polishchuk_MI_2013}.
For the purposes of this investigation, fragment descriptors typically used for organic 
molecules were adapted to serve for materials characterization~\cite{Ruggiu_MI_2010}.
Molecular systems can be described as graphs whose vertices correspond to atoms and edges to chemical bonds. 
In this representation, fragment descriptors characterize subgraphs of the full 3D molecular network.
Any molecular graph invariant can be uniquely represented as a linear combination of fragment descriptors.  
They offer several advantages over other types of chemical descriptors~\cite{Todeschini_MolecularDescriptors_2000}, 
including simplicity of calculation, storage, and interpretation~\cite{Baskin_FragQSAR_2008,Varnek_JCAMD_2005,Baskin_CCHTS_2008}. 
However, they also come with a few disadvantages.  
Models built with fragment descriptors perform poorly when presented with new fragments for which they were not trained.  
Additionally, typical fragments are constructed solely with information of the individual atomic symbols 
(\textit{e.g.}, C, N, Na).
Such a limited context would be insufficient for modeling the complex chemical interactions within materials. 

Mindful of these constraints, novel fragment descriptors for materials were conceptualized
by differentiating atoms not 
by their symbols but by a plethora of well-tabulated chemical and physical properties~\cite{Lide_CRC_2004}.
Descriptor features comprise of various combinations of these atomic properties. 
From this perspective, materials can be thought of as ``colored'' graphs, with vertices decorated according 
to the nature of the atoms they represent~\cite{Varnek_CCADD_2008}.
Partitions of these graphs form \underline{P}roperty-\underline{L}abeled \underline{M}aterials \underline{F}ragments (\PLMF). 

Figure~\ref{figure1} shows the scheme for constructing {\PLMF}s.
Given a crystal structure, the first step is to determine the atomic connectivity within it.
In general, atomic connectivity is not a trivial property to determine within materials.  
Not only must the potential bonding distances among atoms be considered, but also whether 
the topology of nearby atoms allows for bonding.
Therefore, a computational geometry approach is employed to partition the crystal structure (Figure~\ref{figure1}(a)) 
into atom-centered Voronoi-Dirichlet 
polyhedra~\cite{Bowyer_CJ_1981,Blatov_CR_2004,Blatov_ActaCristA_1995,Carlucci_ChemRev_2014} (Figure~\ref{figure1}(b)).
This partitioning scheme was found to be invaluable in the topological analysis of \underline{m}etal 
\underline{o}rganic \underline{f}rameworks (\MOF), molecules, and inorganic crystals~\cite{Baburin_JSSC_2005,Zolotarev_CGD_2014}.
Connectivity between atoms is established by satisfying two criteria:  
\textit{(i)} the atoms must share a Voronoi face (perpendicular bisector between neighboring atoms), and 
\textit{(ii)} the interatomic distance must be shorter than the sum of the Cordero covalent 
radii~\cite{Cordero_DT_2008} to within a 0.25 \text{\AA} tolerance. 
Here, only strong interatomic interactions are modeled, such as covalent, ionic, and metallic 
bonding, ignoring van der Waals interactions.
Due to the ambiguity within materials, the bond order (single/double/triple bond classification) is not considered.
Taken together, the Voronoi centers that share a Voronoi face and are within the sum of their 
covalent radii form a three-dimensional graph defining the connectivity within the material.

In the final steps of the \PLMF\ construction, the full graph and 
corresponding adjacency matrix (Figure~\ref{figure1}(c)) are constructed from the total list of connections.
The adjacency matrix $\mathbf{A}$ of a simple graph (material) with $n$ vertices (atoms) is a square 
matrix $\left(n \times n\right)$ with entries $a_{ij}=1$ if atom $i$ is 
connected to atom $j$, and $a_{ij}=0$ otherwise.
This adjacency matrix reflects the global topology for a given system, including interatomic bonds and contacts within the crystal.  
The full graph is partitioned into smaller subgraphs, corresponding to individual fragments (Figure~\ref{figure1}(d)).  
While there are several subgraphs to consider in general, the length $l$ is restricted to a maximum of three, 
where $l$ is the largest number of consecutive, non-repetitive edges in the subgraph.
This restriction serves to curb the complexity of the final descriptor vector.  
In particular, there are two types of fragments.
Path fragments are subgraphs of at most $l=3$ that encode any linear strand of up to four atoms.  
Only the shortest paths between atoms are considered.
Circular fragments are subgraphs of $l=2$ that encode the first shell of nearest neighbor atoms.
In this context, circular fragments represent coordination polyhedra, or clusters of atoms with 
anion/cation centers each surrounded by a set of its respective counter ion.
Coordination polyhedra are used extensively in crystallography and mineralogy~\cite{pauling_bond}.

The {\PLMF}s are differentiated by local (standard atomic/elemental) reference properties~\cite{Lide_CRC_2004},
which include:
\textit{(i)} general properties: 
the Mendeleev group and period numbers $\left(g_\sP,~p_\sP\right)$,                     
number of valence electrons $\left(N_\sV\right)$;  
\textit{(ii)} measured properties~\cite{Lide_CRC_2004}: 
atomic mass $\left(m_\satom\right)$,                  
electron affinity $\left(EA\right)$,           
thermal conductivity $\left(\lambda\right)$, 
heat capacity $\left(C\right)$, 
enthalpies of atomization $\left(\Delta H_{\mathrm{at}}\right)$, 
fusion $\left(\Delta H_\sfusion\right)$, and
vaporization $\left(\Delta H_\svapor\right)$,
first three ionization potentials $\left(IP_{1,2,3}\right)$; and    
\textit{(iii)} derived properties: 
effective atomic charge $\left(Z_\seff\right)$,    
molar volume $\left(V_\smolar\right)$,             
chemical hardness $\left(\eta\right)$~\cite{Parr_JACS_1983,Lide_CRC_2004}, 
covalent $\left(r_\scov\right)$~\cite{Cordero_DT_2008}, 
absolute~\cite{Ghosh_IJMS_2002}, and                        
van der Waals radii~\cite{Lide_CRC_2004},                   
electronegativity $\left(\chi\right)$,                      
and
polarizability $\left(\alpha_\sP\right)$.  
Pairs of properties are included in the form of their multiplication and ratio, 
as well as the property value divided by the atomic connectivity 
(number of neighbors in the adjacency matrix). 
For every property scheme $\mathbf{q}$, the following quantities are also considered:
minimum $\left(\min(\mathbf{q})\right)$, 
maximum $\left(\max(\mathbf{q})\right)$, 
total sum $\left(\sum \mathbf{q}\right)$, 
average $\left(\avg (\mathbf{q}) \right)$, and 
standard deviation $\left(\std(\mathbf{q})\right)$ of 
$\mathbf{q}$ among the atoms in the material.

To incorporate information about shape, size, and symmetry of the crystal unit cell, 
the following crystal-wide properties are incorporated: 
lattice parameters ($a$, $b$, $c$), their ratios ($a/b$, $b/c$, $a/c$), angles ($\alpha$, $\beta$, $\gamma$), 
density, volume, volume per atom, number of atoms, number of species (atom types), 
lattice type, point group, and space group.  

All aforementioned descriptors (fragment-based and crystal-wide) can be concatenated together to represent each
material uniquely.
After filtering out low variance ($<0.001$) and highly correlated $\left(r^{2}\!>\!0.95\right)$ features, 
the final feature vector captures 2,494 total descriptors.

Descriptor construction is inspired by the topological charge indices~\cite{Galvez_JCICS_1995}
and the Kier-Hall 
electro-topological state indices~\cite{Kier_Electrotopological_1999,Hall_QSAR_1991}.
Let $\mathbf{M}$ be the matrix obtained by multiplying the adjacency 
matrix $\mathbf{A}$ by the reciprocal square distance matrix $\mathbf{D}$ $\left(D_{ij}=1/r_{i,j}^{2}\right)$:
\[
\mathbf{M}=\mathbf{A} \cdot \mathbf{D}.
\]
The matrix $\mathbf{M}$, called the Galvez matrix, is a square $n \times n$ matrix, 
where $n$ is the number of atoms in the unit cell.  
From $\mathbf{M}$, descriptors of reference property $\mathbf{q}$ are calculated as
\[
T^{\mathrm{E}}=\sum_{i=1}^{n-1}\sum_{j=i+1}^{n}\left|q_{i}-q_{j}\right|M_{ij}
\]
and
\[
T_{\sbond}^{\mathrm{E}}=\sum_{\{i,j\}\in\mathrm{bonds}}\left|q_{i}-q_{j}\right|M_{ij},
\]
where the first set of indices count over all pairs of atoms and the second 
is restricted to all pairs $i,j$ of bonded atoms. 

\begin{figure*}[t!]
  \includegraphics[width=\textwidth]{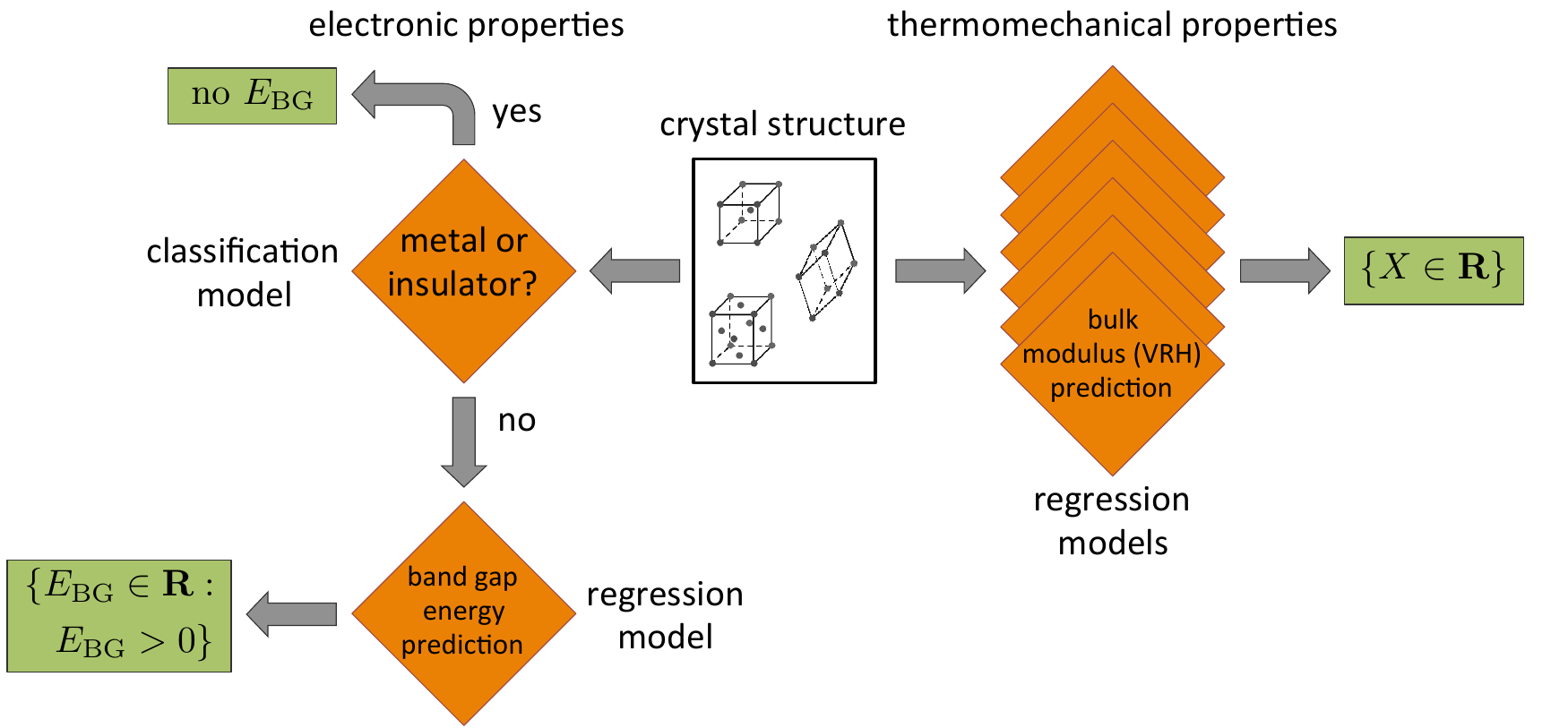}
  \caption{\small
    \textbf{Outline of the modeling work-flow.}
    \ML\ models are represented by orange diamonds. Target properties predicted by these models are highlighted in green.}
  \label{figure2}
\end{figure*}
\ \\

\noindent\textbf{Quantitative Materials Structure-Property Relationship modeling.}
In training the models, the same \ML\ method and descriptors are employed without any hand tuning or variable selection.
Specifically, models are constructed using \underline{g}radient \underline{b}oosting \underline{d}ecision \underline{t}ree 
(\GBDT) technique~\cite{Friedman_AnnStat_2001}.
All models were validated through $y$-randomization (label scrambling).  
Five-fold cross validation is used to assess how well each model will generalize to an independent dataset.
Hyperparameters are determined with grid searches on the training set and 10-fold cross validation.

The \underline{g}radient \underline{b}oosting \underline{d}ecision \underline{t}rees (\GBDT) method \cite{Friedman_AnnStat_2001} 
evolved from the application of boosting 
methods~\cite{gbm} to regression trees~\cite{Loh_ISR_2014}.
The boosting method is based on the observation that finding many weakly accurate 
prediction rules can be a lot easier than finding a single, highly accurate rule~\cite{Schapire_ML_1990}.
The boosting algorithm calls this ``weak'' learner repeatedly, at each stage feeding it 
a different subset of the training examples. 
Each time it is called, the weak learner generates a new weak prediction rule.
After many iterations, the boosting algorithm combines these weak rules into 
a single prediction rule aiming to be much more accurate than any single weak rule.

The \GBDT\ approach is an additive model of the following form:
\[
F(\mathbf{x};\{\gamma_{m},\mathbf{a}\}_{1}^{M})=\sum_{m=1}^{M}\gamma_{m} h_{m}(\mathbf{x};\mathbf{a}_{m}),
\]
where $h_{m}(\mathbf{x};\mathbf{a}_{m})$ are the weak learners (decision trees in this case) 
characterized by parameters
$\mathbf{a}_{m}$, and $M$ is the total 
count of decision trees obtained through boosting.

It builds the additive model in a forward stage-wise fashion:
\[
F_m(\mathbf{x})=F_{m-1}(\mathbf{x})+\gamma_{m} h_{m}(\mathbf{x};\mathbf{a}_{m}).
\]
At each stage $\left(m=1,2,\ldots,M\right)$, $\gamma_{m}$ and $\mathbf{a}_{m}$ are chosen to minimize the loss function 
$f_L$ given the current model $F_{m-1}(x_{i})$ for all data points (count $N$),
\[
\left(\gamma_{m},\mathbf{a}_{m}\right)=\argmin_{\gamma,\mathbf{a}} \sum_{i=1}^{N}
f_{L} \left[y_{i},F_{m-1}\left(\mathbf{x_{i}}\right)+\gamma h\left(\mathbf{x}_{i};\mathbf{a}\right)\right].
\]
Gradient boosting attempts to solve this minimization problem numerically via steepest descent. 
The steepest descent direction is the negative gradient of the loss function 
evaluated at the current model $F_{m-1}$, where the step length is chosen using line search.

An important practical task is to quantify variable importance. 
Feature selection in decision tree ensembles cannot differentiate between primary 
effects and effects caused by interactions between variables.
Therefore, unlike regression coefficients, a direct comparison of captured effects is prohibited.
For this purpose, variable influence is quantified in the 
following way~\cite{Friedman_AnnStat_2001}.
Let us define the influence of variable $j$ in a single tree $h$. 
Consider that the tree has $l$ splits and therefore $l-1$ levels. 
This gives rise to the definition of the variable influence,
\[
K_{j}^2(h)=\sum_{i=1}^{l-1} I_{i}^{2} \mathbbm{1}\left(x_{i}=j\right),
\]
where $I_{i}^{2}$ is the empirical squared improvement resulting from this split,
and $\mathbbm{1}$ is the indicator function.
Here, $\mathbbm{1}$ has a value of one if the split at node $x_{i}$ is on variable $j$, and
zero otherwise, 
\textit{i.e.}, it measures the number of times a variable $j$ is selected for splitting. 
To obtain the overall influence of variable $j$ in the ensemble of decision trees (count $M$), 
it is averaged over all trees,
\[
K_{j}^{2}={M}^{-1} \sum_{m=1}^{M} K_{j}^{2} (h_{m}).
\]
The influences $K_{j}^{2}$ are normalized so that they add to one. 
Influences capture the importance of the variable, but 
not the direction of the response (positive or negative). 

\begin{figure*}[t!]
  \includegraphics[width=\textwidth]{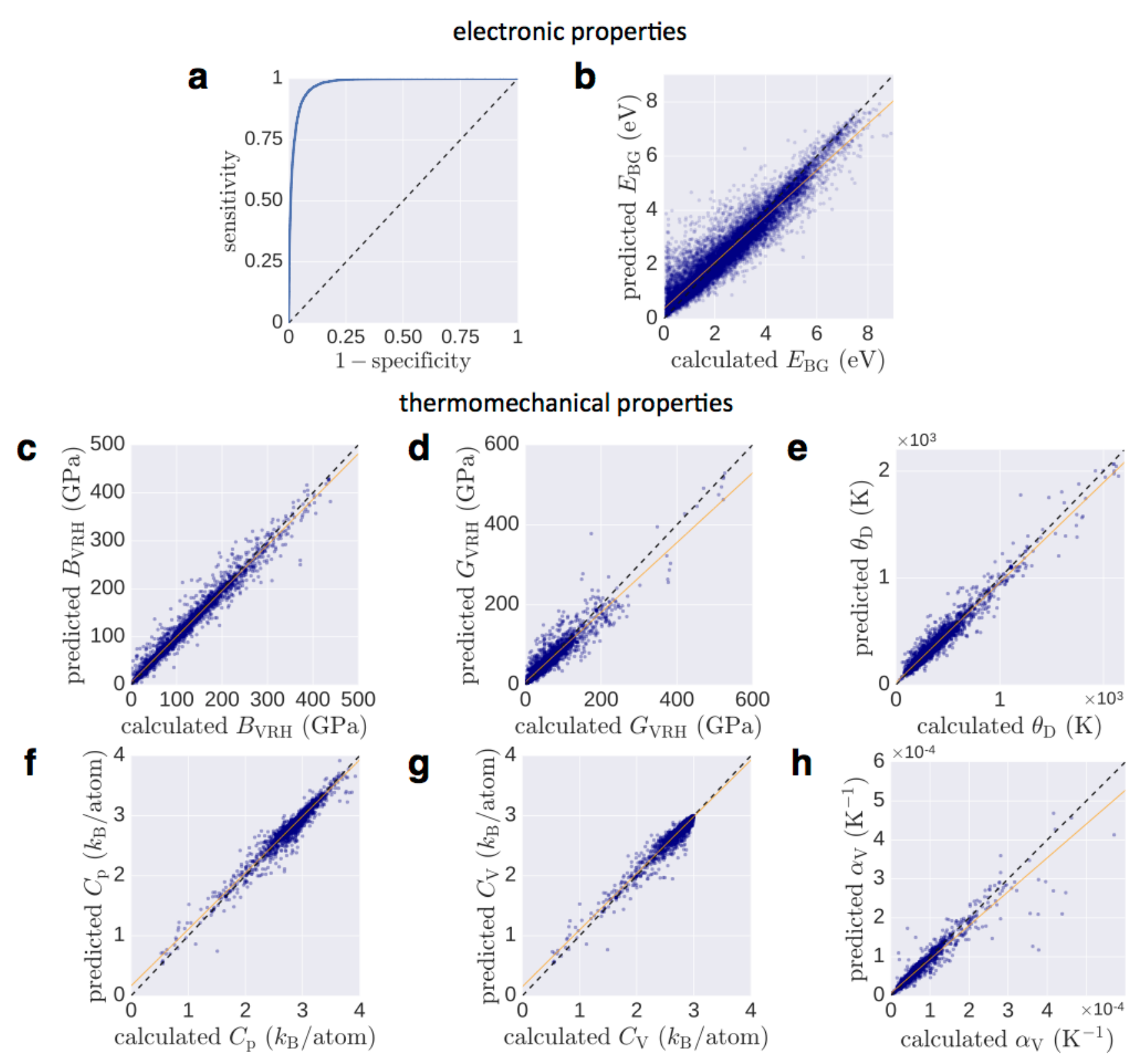}
  \caption{\small 
    \textbf{Five-fold cross validation plots for the eight \ML\ models predicting electronic and thermomechanical properties.}
    \textbf{(a)} \underline{R}eceiver \underline{o}perating \underline{c}haracteristic (\ROC) curve for the classification \ML\ model.  
    \textbf{(b)}-\textbf{(h)} Predicted \textit{vs.} calculated values
    for the regression \ML\ models:
    \textbf{(b)} band gap energy $\left(E_{\scriptstyle \mathrm{BG}}\right)$, 
    \textbf{(c)} bulk modulus $\left(B_{\scriptstyle \mathrm{VRH}}\right)$, 
    \textbf{(d)} shear modulus $\left(G_{\scriptstyle \mathrm{VRH}}\right)$, 
    \textbf{(e)} Debye temperature $\left(\theta_{\scriptstyle \mathrm{D}}\right)$, 
    {\textbf{(f)} heat capacity at constant pressure $\left(C_{\scriptstyle \mathrm{P}}\right)$, 
    \textbf{(g)} heat capacity at constant volume $\left(C_{\scriptstyle \mathrm{V}}\right)$}, and 
    \textbf{(h)} thermal expansion coefficient $\left(\alpha_{\scriptstyle \mathrm{V}}\right)$. 
  }
  \label{figure3}
\end{figure*}

\noindent \textbf{Integrated modeling work-flow.}
Eight predictive models are developed in this work, including:
a binary classification model that predicts if a material is a metal or an insulator
and seven regression models that predict:
the band gap energy $\left(E_\sBG\right)$ for insulators,
bulk modulus $\left(B_\sVRH\right)$,
shear modulus $\left(G_\sVRH\right)$,
Debye temperature $\left(\theta_\sD\right)$,
heat capacity at constant pressure $\left(C_\sp\right)$,
heat capacity at constant volume $\left(C_\sV\right)$, and
thermal expansion coefficient $\left(\alpha_\sV\right)$.

Figure~\ref{figure2} shows the overall application work-flow.
A novel candidate material is first classified as a metal or an insulator.
If the material is classified as an insulator, $E_\sBG$ is predicted, 
while classification as a metal implies that the material has no $E_\sBG$. 
The six thermomechanical properties are then predicted independent of the material's metal/insulator classification.
The integrated modeling work-flow has been implemented as a web application at 
\href{http://aflow.org/aflow-ml}{aflow.org/aflow-ml}, 
requiring only the atomic species and positions as input for predictions.

While all three models were trained independently, the accuracy of the 
$E_\sBG$ regression model is inherently dependent on the accuracy of the metal/insulator classification model
in this work-flow.
However, the high accuracy of the metal/insulator classification model suggests this not to be a practical concern.

\section{Results}
\noindent \textbf{Model generalizability.} 
One technique for assessing model quality is five-fold cross validation, which gauges how well
the model is expected to generalize to an independent dataset.
For each model, the scheme involves randomly partitioning the set into five groups and predicting the value of 
each material in one subset while training the model on the other four subsets.
Hence, each subset has the opportunity to play the role of the ``test set''.
Furthermore, any observed deviations in the predictions are addressed.
For further analysis, all predicted and calculated results are available in Supplemental Information.

The accuracy of the metal/insulator classifier is reported as the 
\underline{a}rea \underline{u}nder the \underline{c}urve (\AUC) 
of the \underline{r}eceiver \underline{o}perating \underline{c}haracteristic (\ROC) plot (Figure~\ref{figure3}(a)).
The \ROC\ curve illustrates the model's ability to differentiate between metallic and insulating input materials. 
It plots the prediction rate for insulators (correctly \textit{vs.} incorrectly predicted) throughout the 
full spectrum of possible prediction thresholds. 
An area of 1.0 represents a perfect test, while an area of 0.5 characterizes a random guess (the dashed line).
The model shows excellent external predictive power with the \AUC\ at 0.98, 
an insulator-prediction success rate (sensitivity) of 0.95, 
a metal-prediction success rate (specificity) of 0.92,
and an overall classification rate (\CCR) of 0.93.
For the complete set of \electronicTotal\ materials, this corresponds to
2,103 misclassified materials, including 1,359 misclassified metals and 744 misclassified insulators.
Evidently, the model exhibits positive bias toward predicting insulators, where bias refers to whether a 
\ML\ model tends to over- or under-estimate the predicted property. 
This low false-metal rate is fortunate as the model is unlikely to 
misclassify a novel, potentially interesting semiconductor as a metal.
Overall, the metal classification model is robust enough to handle the full complexity of the periodic table.
 
\begin{table}[t!]
\begin{tabular}{|c||c|c|c|}
\hline
property & \RMSE\ & \MAE\ & $r^{2}$ \\
\hline \hline
$E_\sBG$                                & 0.51 eV                                     & 0.35 eV                                 & 0.90 \\
\hline
$B_\sVRH$                               & 14.25 GPa                                   & 8.68 GPa                                & 0.97 \\
\hline
$G_\sVRH$                               & 18.43 GPa                                   & 10.62 GPa                               & 0.88 \\
\hline
$\theta_\sD$                            & 56.97 K                                     & 35.86 K                               & 0.95 \\
\hline
$C_\sp$                                 & {0.09 $k_\sB$/atom}                                & {0.05 $k_\sB$/atom}                            & {0.95} \\
\hline
$C_\sV$                                 & {0.07 $k_\sB$/atom}                                & {0.04 $k_\sB$/atom}                            & {0.95} \\
\hline
$\alpha_\sV$                            & $1.47 \times 10^{-5}$ K$^{-1}$               & $5.69 \times 10^{-6}$ K$^{-1}$           & 0.91 \\
\botrule
\end{tabular}
\caption{\small
  Statistical summary of the \textit{five-fold cross-validated predictions} for the seven regression models (Figure~\ref{figure3}).
}
\label{table1}
\end{table}

The results of the five-fold cross validation analysis for the band gap energy $\left(E_\sBG\right)$ regression model
are plotted in Figure~\ref{figure3}(b).
Additionally, a statistical profile of these predictions, along with that of the six thermomechanical regression models,
is provided in Table~\ref{table1}, which includes metrics such as
the \underline{r}oot-\underline{m}ean-\underline{s}quare \underline{e}rror (\RMSE),
\underline{m}ean \underline{a}bsolute \underline{e}rror (\MAE), and coefficient of determination $\left(r^2\right)$.
Similar to the classification model, the $E_\sBG$ model exhibits a positive predictive bias.  
The biggest errors come from materials with narrow band gaps, 
\textit{i.e.}, the scatter in the lower left corner in Figure~\ref{figure3}(b). 
These materials predominantly include complex fluorides and nitrides. 
N$_{2}$H$_{6}$Cl$_{2}$ (\ICSD\ \#23145) 
exhibits the worst prediction accuracy with \underline{s}igned \underline{e}rror SE = 3.78 eV~\cite{Donohue_JCP_1947}.
The most underestimated materials are HCN (\ICSD\ \#76419) and, respectively 
N$_{2}$H$_{6}$Cl$_{2}$ (\ICSD\ \#240903) with SE = -2.67 and -3.19 eV~\cite{Dulmage_ActaCrist_1951,Kruszynski_ActaCristE_2007}, respectively. 
This is not surprising considering that all three are molecular crystals.
Such systems are anomalies in the \ICSD, and fit better in other databases, such as
the Cambridge Structural Database~\cite{Groom_CSD_2016}.
Overall, 10,762 materials are predicted within 25\% accuracy of calculated values, 
whereas 824 systems have errors over 1 eV.

Figures~\ref{figure3}(c-h) and Table~\ref{table1} showcase the results of the five-fold cross validation analysis
for the six thermomechanical regression models.
For both bulk $\left(B_\sVRH\right)$ and shear $\left(G_\sVRH\right)$ moduli, 
over 85\% of materials are predicted within 20 GPa of their calculated values.
The remaining models also demonstrate high accuracy, with 
at least 90\% of the full training set $\left(>2,546~\mathrm{systems}\right)$ 
predicted to within 25\% of the calculated values.
Significant outliers in predictions of the bulk modulus include 
graphite (\ICSD\ \#187640, SE = 100 GPa, likely
due to extreme anisotropy) and two theoretical high-pressure boron nitrides (\ICSD\ \#162873 and \#162874, 
under-predicted by over 110 GPa)~\cite{Lian_JCP_2013,Doll_PRB_2008}.
Other theoretical systems are ill-predicted throughout the six properties, including
ZN (\ICSD\ \#161885), CN$_{2}$ (\ICSD\ \#247676), C$_{3}$N$_{4}$ (\ICSD\ \#151782),
and CH (\ICSD\ \#187642)~\cite{EscorciaSalas_MJ_2008,Li_PCCP_2012,Marques_PRB_2004,Lian_JCP_2013}.
Predictions for the $G_\sVRH$, Debye temperature $\left(\theta_\sD\right)$, and thermal expansion coefficient
$\left(\alpha_\sV\right)$
tend to be slightly underestimated, particularly for higher calculated values.
Additionally, mild scattering can be seen for $\theta_\sD$ and 
$\alpha_\sV$, but not enough to have a significant
impact on the error or correlation metrics.

Despite minimal deviations, both \RMSE\ and \MAE\ are within 4\% of the ranges covered for each property, 
and the predictions demonstrate excellent correlation with the calculated properties.
{Note the tight clustering of points just below 3 $k_\sB$/atom for the heat
capacity at constant volume $\left(C_\sV\right)$.
This is due to $C_\sV$ saturation in accordance with the Dulong-Petit law occurring at or below
300 K for many compounds.}

\begin{figure}[t!]
  \center
  \includegraphics[width=0.48\textwidth]{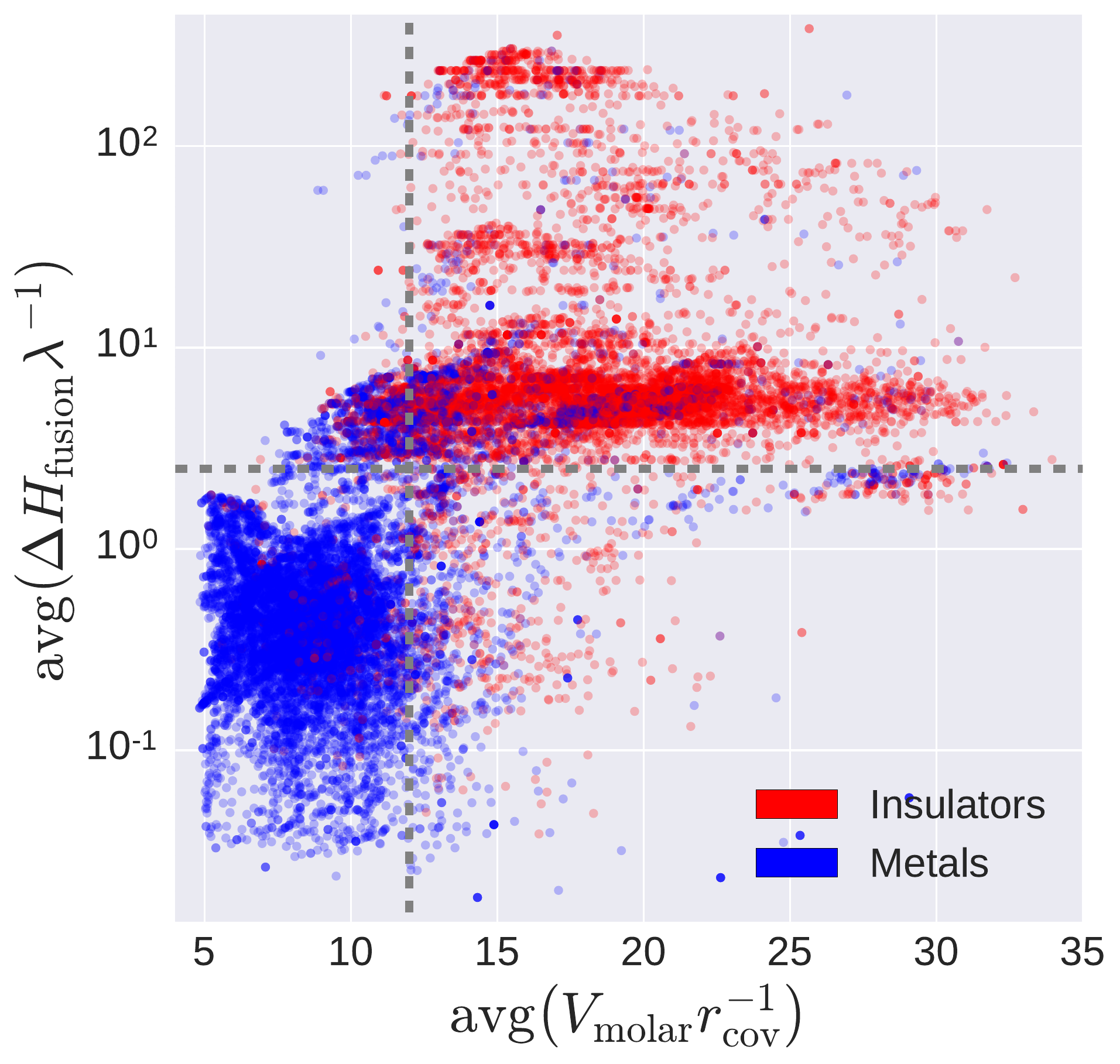}
  \caption{\small 
    \textbf{Semi-log plot of the full dataset (\electronicTotal\ unique materials) in the dual-descriptor 
    space} of $\avg\left(\Delta H_{\scriptstyle \mathrm{fusion}}\lambda^{-1}\right)$ and 
    $\avg\left(V_{\scriptstyle \mathrm{molar}}r_{\scriptstyle \mathrm{cov}}^{-1}\right)$.
    Insulators and metals are colored in red and blue, respectively.
  }
  \label{figure4}
\end{figure}

\begin{figure*}[t!]
  \includegraphics[width=\linewidth]{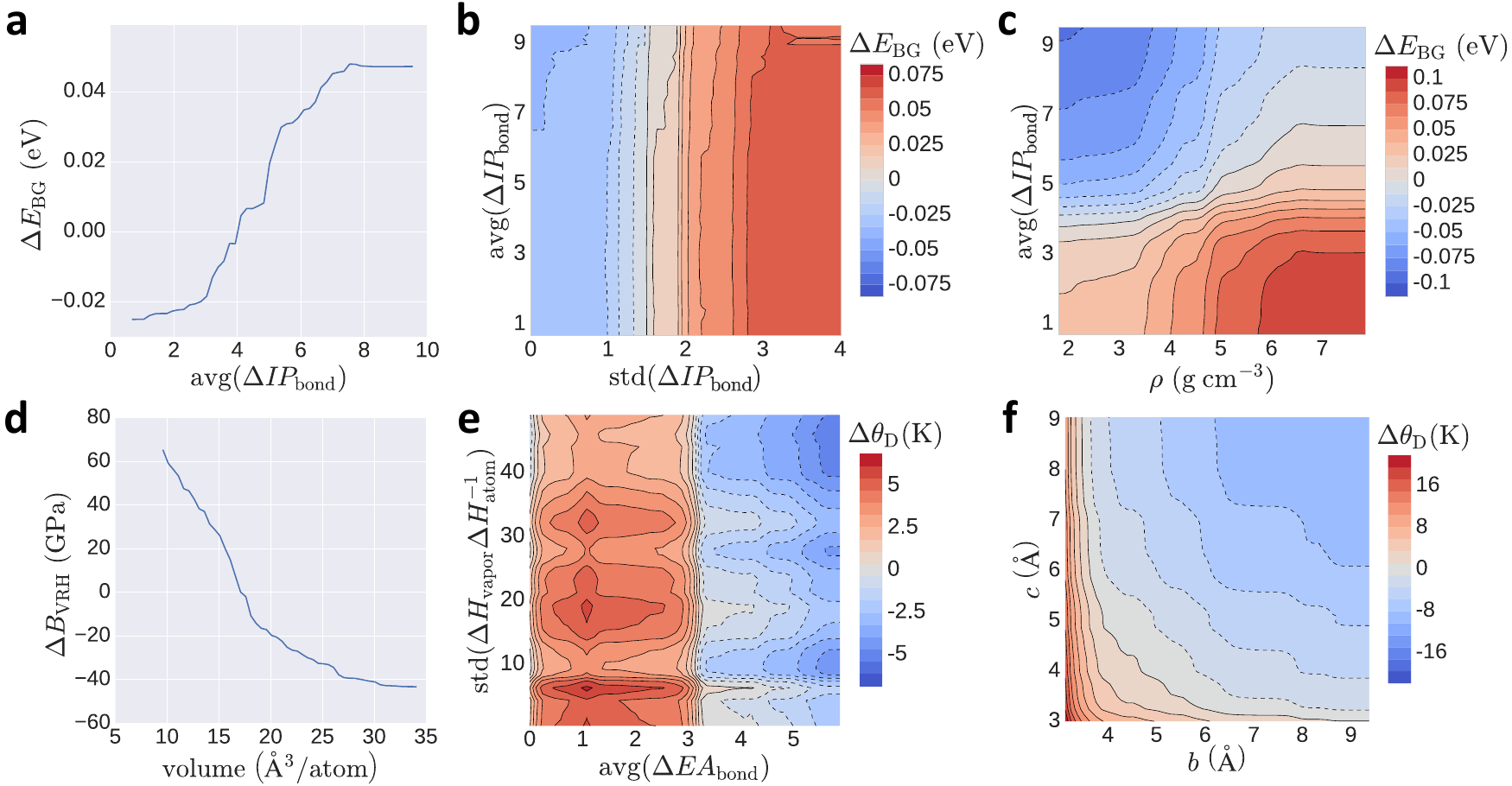}%
  \caption{\small
    \textbf{Partial dependence plots of the $E_{\scriptstyle \mathrm{BG}}$, $B_{\scriptstyle \mathrm{VRH}}$, and 
    $\theta_{\scriptstyle \mathrm{D}}$ models.}
    \textbf{(a)} Partial dependence of $E_{\scriptstyle \mathrm{BG}}$ on the $\avg\left(\Delta IP_{\mathrm{bond}}\right)$
    descriptor. 
    For $E_{\scriptstyle \mathrm{BG}}$, 
    the 2D interaction between $\std\left(\Delta IP_{\mathrm{bond}}\right)$ and $\avg\left(\Delta IP_{\mathrm{bond}}\right)$
    and between $\rho$ (density) and $\avg\left(\Delta IP_{\mathrm{bond}}\right)$ are illustrated in panels 
    \textbf{(b)} and \textbf{(c)}, respectively. 
    \textbf{(d)} Partial dependence of the $B_{\scriptstyle \mathrm{VRH}}$ on the crystal volume per atom descriptor.
    For $\theta_{\scriptstyle \mathrm{D}}$, 
    the 2D interaction between 
    $\avg\left(\Delta EA_{\scriptstyle \mathrm{bond}}\right)$ and 
    $\std\left(\Delta H_{\scriptstyle \mathrm{vapor}} \Delta H_{\scriptstyle \mathrm{atom}}^{-1}\right)$ 
    and between 
    crystal lattice parameters $b$ and $c$ are illustrated
    in panels \textbf{(e)} and \textbf{(f)}, respectively. 
  }
  \label{figure5}
\end{figure*}

\noindent \textbf{Model interpretation.}
Model interpretation is of paramount importance in any \ML\ study.
The significance of each descriptor is determined in order to gain insight into
structural features that impact molecular properties of interest.
Interpretability is a strong advantage of decision tree methods, particularly with the \GBDT\ approach.
One can quantify the predictive power of a specific descriptor by analyzing the reduction 
of the \RMSE\ at each node of the tree.

Partial dependence plots offer yet another opportunity for \GBDT\ model interpretation.
Similar to the descriptor significance analysis, partial dependence resolves the 
effect of a variable (descriptor) on a property, but only after marginalizing over all other 
explanatory variables~\cite{Hastie_StatLearn_2001}.
The effect is quantified by the change of that property as relevant descriptors are varied.
The plots themselves highlight the most important interactions among relevant descriptors 
as well as between properties and their corresponding descriptors.

While only the most important descriptors are highlighted and discussed, 
an exhaustive list of relevant descriptors and their relative contributions can be found in the Supplementary Information.

For the metal/insulator classification model, the descriptor significance analysis 
shows that two descriptors have the highest importance (equally), namely 
$\avg\left(\Delta H_\sfusion\lambda^{-1}\right)$ and 
$\avg\left(V_\smolar r_\scov^{-1}\right)$.
$\avg\left(\Delta H_\sfusion\lambda^{-1}\right)$ is the ratio between the 
fusion enthalpy $\left(\Delta H_\sfusion\right)$ 
and the thermal conductivity $\left(\lambda\right)$ averaged over all atoms in the material, and 
$\avg\left(V_\smolar r_\scov^{-1}\right)$ is the ratio between the 
molar volume $\left(V_\smolar\right)$ 
and the covalent radius $\left(r_\scov\right)$ averaged over all atoms in the material. 
Both descriptors are simple node-specific features.
The presence of these two prominent descriptors accounts for the high accuracy of the classification model.

Figure~\ref{figure4} shows the projection of the full dataset onto the dual-descriptor space of 
$\avg\left(\Delta H_\sfusion\lambda^{-1}\right)$ and $\avg\left(V_\smolar r_\scov^{-1}\right)$.  
In this 2D space, metals and insulators are substantially partitioned.
To further resolve this separation, the plot is split into four quadrants 
(see dashed lines) with an origin approximately at 
\[
\avg\left(V_\smolar r_\scov^{-1}\right)=11,\quad \avg\left(\Delta H_\sfusion\lambda^{-1}\right)=2.
\]
Insulators are predominately located in quadrant I.
There are several clusters (one large and several small) parallel to the $x$-axis.
Metals occupy a compact square block in quadrant III within intervals 
$5<\avg\left(V_\smolar r_\scov^{-1}\right)<12$ and $0.02<\avg\left(\Delta H_\sfusion\lambda^{-1}\right)<2$.
Quadrant II is mostly empty with a few materials scattered about the origin.
In the remaining quadrant (IV), materials have mixed character.

Analysis of the projection shown in Figure~\ref{figure4} suggests a simple heuristic rule:
all materials within quadrant I are classified as insulators $\left(E_\sBG\!>\!0\right)$, 
and all materials outside of this quadrant are metals. 
Remarkably, this unsupervised projection approach achieves a very high 
classification accuracy of 86\% for the entire dataset of \electronicTotal\ materials.  
The model misclassifies only 3,621 materials:
2,414 are incorrectly predicted as insulators and 1,207 are incorrectly predicted as metals.  
This example illustrates how careful model analysis of the most significant descriptors
can yield simple heuristic rules for materials design.

The regression model for the band gap energy $\left(E_\sBG\right)$ is more complex.
There are a number of descriptors in the model with comparable contributions, 
and thus, all individual contributions are small.
This is expected as a number of conditions can affect $E_\sBG$.
The most important are $\avg\left(\chi Z_{\mathrm{eff}}^{-1}\right)$ and $\avg\left(C \lambda^{-1}\right)$ with 
significance scores of 0.075 and 0.071, 
respectively, where $\chi$ is the electronegativity, $Z_{\mathrm{eff}}$ is the effective nuclear charge, 
$C$ is the specific heat capacity, and $\lambda$ is the thermal conductivity of each atom.  

Figure~\ref{figure5} shows partial dependence plots focusing on $\avg\left(\Delta IP_{\sbond}\right)$ as an example.
It is derived from edge fragments of bonded atoms $\left(l=1\right)$ and defined as an absolute difference in 
ionization potentials averaged over the material.
In other words, it is a measure of bond polarity, similar to electronegativity.
Figure~\ref{figure5}(a) shows a steady monotonic increase in $\Delta E_\sBG$ for larger 
values of $\avg\left(\Delta IP_{\sbond}\right)$. 
The effect is small, but captures an expected physical principle:
polar inorganic materials (\textit{e.g.}, oxides, fluorides) tend to have larger $E_\sBG$.

\begin{figure*}
  \includegraphics[width=\linewidth]{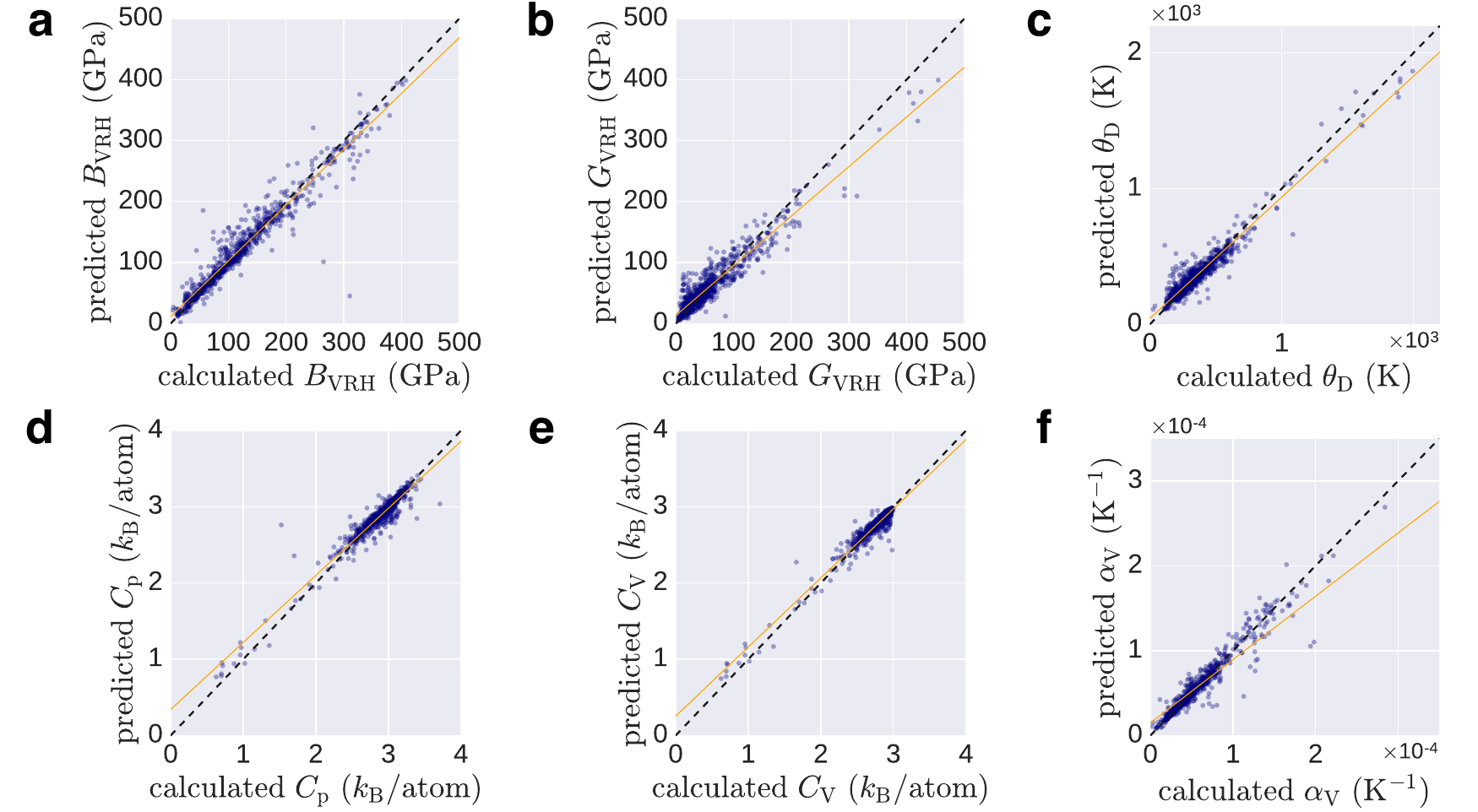}%
  \caption{\small
      \textbf{Model performance evaluation for the six \ML\ models predicting thermomechanical properties 
      of \thermoTestTotal\ newly characterized materials.}
      Predicted \textit{vs.} calculated values for the regression \ML\ models:
      \textbf{(a)} bulk modulus $\left(B_{\scriptstyle \mathrm{VRH}}\right)$,
      \textbf{(b)} shear modulus $\left(G_{\scriptstyle \mathrm{VRH}}\right)$,
      \textbf{(c)} Debye temperature $\left(\theta_{\scriptstyle \mathrm{D}}\right)$,
      {\textbf{(d)} heat capacity at constant pressure $\left(C_{\scriptstyle \mathrm{P}}\right)$,
      \textbf{(e)} heat capacity at constant volume $\left(C_{\scriptstyle \mathrm{V}}\right)$}, and
      \textbf{(f)} thermal expansion coefficient $\left(\alpha_{\scriptstyle \mathrm{V}}\right)$.
  }
  \label{figure6}
\end{figure*}

Given the number of significant interactions involved with this phenomenon, 
tailoring $E_\sBG$ involves the 
optimization of a highly non-convex, multidimensional object.
Figure~\ref{figure5}(b) illustrates a 2D slice of this object as
$\std\left(\Delta IP_{\sbond}\right)$ and $\avg\left(\Delta IP_{\sbond}\right)$ vary simultaneously.
Like $\avg\left(\Delta IP_{\sbond}\right)$,
$\std\left(\Delta IP_{\sbond}\right)$ is the standard deviation of the set of absolute differences in $IP$ among
all bonded atoms.
In the context of these two variables, $E_\sBG$ responds to deviations in $\Delta IP_{\sbond}$ 
among the set of bonded atoms, but remains constant across shifts in $\avg\left(\Delta IP_{\sbond}\right)$.
This suggests an opportunity to tune $E_\sBG$ by considering another composition that varies the deviations among bond polarities.
Alternatively, a desired $E_\sBG$ can be maintained
by considering another composition that preserves the deviations among bond polarities, even as the overall average 
shifts.
Similarly, Figure~\ref{figure5}(c) shows the partial dependence on both 
the density $\left(\rho\right)$ and $\avg\left(\Delta IP_{\sbond}\right)$.
Contrary to the previous trend, larger $\avg\left(\Delta IP_{\sbond}\right)$
values correlate with smaller $E_\sBG$, particularly for low density structures.
Materials with higher density and lower $\avg\left(\Delta IP_{\sbond}\right)$ tend to have higher $E_\sBG$.
Considering the elevated response (compared to Figure~\ref{figure5}(b)), the inverse correlation of $E_\sBG$ with the average
bond polarity in the context of density suggests an even more effective means of tuning $E_\sBG$. 

A descriptor analysis of the thermomechanical property models reveals the importance of
one descriptor in particular, the volume per atom of the crystal.
This conclusion certainly resonates with the nature of these properties, as they generally correlate
with bond strength~\cite{Toher_AEL_2016}.
Figure~\ref{figure5}(d) exemplifies such a relationship, which shows
the partial dependence plot of the bulk modulus $\left(B_\sVRH\right)$ on the volume per atom.
Tightly bound atoms are generally indicative of stronger bonds.
As the interatomic distance increases, properties like $B_\sVRH$ generally reduce.

\begin{table}
\begin{tabular}{|c||c|c|c|}
\hline
property & \RMSE\ & \MAE\ & $r^{2}$ \\
\hline \hline
$B_\sVRH$                           & 21.13 GPa                               & 12.00 GPa                               & 0.93 \\
\hline
$G_\sVRH$                           & 18.94 GPa                               & 13.31 GPa                               & 0.90 \\
\hline
$\theta_\sD$                        & 64.04 K                                 & 42.92 K                                 & 0.93 \\
\hline
$C_\sp$                             & {0.10 $k_\sB$/atom}                            & {0.06 $k_\sB$/atom}                             & {0.92} \\
\hline
$C_\sV$                             & {0.07 $k_\sB$/atom}                            & {0.05 $k_\sB$/atom}                             & {0.95} \\
\hline
$\alpha_\sV$                        & $1.95 \times 10^{-5}$ K$^{-1}$           & $5.77 \times 10^{-6}$ K$^{-1}$           & 0.76 \\
\botrule
\end{tabular}
\caption{\small
    Statistical summary of the \textit{new predictions} for the six thermomechanical regression models (Figure~\ref{figure6}).
}
\label{table2}
\end{table}

Two of the more interesting dependence plots are also shown in Figure~\ref{figure5}(e-f),
both of which offer opportunities for tuning the Debye temperature ($\theta_\sD$).
Figure~\ref{figure5}(e) illustrates the interactions among two descriptors, 
the absolute difference in electron affinities among bonded atoms
averaged over the material
$\left(\avg\left(\Delta EA_\sbond\right)\right)$, and 
the standard deviation of the set of ratios of the enthalpies of vaporization $\left(\Delta H_\svapor\right)$
and atomization $\left(\Delta H_\satom\right)$ for all atoms in the material
$\left(\std\left(\Delta H_\svapor \Delta H_\satom^{-1}\right)\right)$.
Within these dimensions, two distinct regions emerge of increasing/decreasing $\theta_\sD$ separated by a
sharp division
at about $\avg\left(\Delta EA_\satom\right) = 3$.
Within these partitions, there are clusters of maximum gradient in $\theta_\sD$---peaks within the left
partition and troughs within the right.
The peaks and troughs alternate with varying $\std\left(\Delta H_\svapor \Delta H_\satom^{-1}\right)$.
Although $\std\left(\Delta H_\svapor \Delta H_\satom^{-1}\right)$
is not an immediately intuitive descriptor, the alternating clusters may be a manifestation 
of the periodic nature of $\Delta H_\svapor$ and $\Delta H_\satom$~\cite{webelements_periodicity}.
As for the partitions themselves, 
the extremes of $\avg\left(\Delta EA_\satom\right)$ characterize covalent and ionic materials, as
bonded atoms with similar $EA$ are likely to share electrons, while those
with varying $EA$ prefer to donate/accept electrons.
Considering that $EA$ is also periodic, various opportunities for carefully tuning $\theta_\sD$
should be available.

Finally, Figure~\ref{figure5}(f) shows the partial dependence of $\theta_\sD$ on the lattice parameters $b$ and $c$.
It resolves two notable correlations: 
\textit{(i)} uniformly increasing the cell size of the system decreases $\theta_\sD$, but
\textit{(ii)} elongating the cell ($c/b \gg 1$) increases it.
Again, \textit{(i)} can be attributed to the
inverse relationship between volume per atom and bond strength,
but does little to address \textit{(ii)}.
Nevertheless, the connection between elongated, or layered, systems and the Debye temperature is certainly not
surprising---anisotropy can be leveraged to enhance phonon-related interactions associated with
thermal conductivity~\cite{Minnich_PRB_2015}
and superconductivity~\cite{Shimahara_PRB_2002,Jha_PT_1989,Klein_SSC_1980}.
While the domain of interest is quite narrow,
the impact is substantial, particularly in comparison to that shown in Figure~\ref{figure5}(e).

\noindent \textbf{Model validation.}
While the expected performances of the \ML\ models can be projected through five-fold cross validation,
there is no substitute for validation against an independent dataset.
The \ML\ models for the thermomechanical properties are leveraged to make predictions
for materials previously uncharacterized, and subsequently validated
these predictions via the \AEL-\AGL\ integrated framework~\cite{curtarolo:art96, Toher_AEL_2016}.
Figure~\ref{figure6} illustrates the models' performance on the set of \thermoTestTotal\ additional materials,
with relevant statistics displayed in Table~\ref{table2}.

Comparing with the results of the generalizability analysis shown in Figure~\ref{figure3} and Table~\ref{table1},
the overall errors are consistent with five-fold cross validation. 
Five out of six models have $r^2$ of 0.9 or higher.
However, the $r^2$ value for the thermal expansion coefficient 
$\left(\alpha_\sV\right)$ is lower than forecasted.
The presence of scattering suggests the need for a larger training set---as new, 
much more diverse materials were likely introduced in the test set.
This is not surprising considering the number of variables that can affect thermal expansion~\cite{Figge_APL_2009}.
Otherwise, the accuracy of these predictions confirm the effectiveness of the \PLMF\ representation, 
which is particularly compelling considering:
\textit{(i)} the limited diversity training dataset (only about 11\% as large as that available for
predicting the electronic properties), and
\textit{(ii)} the relative size of the test set (over a quarter the size of the training set).

In the case of the bulk modulus $\left(B_\sVRH\right)$, 665 systems (86\% of test set) are predicted within 25\%
of calculated values.
Only the predictions of four materials, Bi (\ICSD\ \#51674), PrN (\ICSD\ \#168643),
Mg$_{3}$Sm (\ICSD\ \#104868), and ZrN (\ICSD\ \#161885), deviate beyond 100 GPa from calculated values. 
Bi is a high-pressure phase (Bi-III) with a caged, zeolite-like structure~\cite{McMahon_BiIII_2001}.
The structures of zirconium nitride (wurtzite phase) and praseodymium nitride (B3 phase) were hypothesized and
investigated via \DFT\ calculations~\cite{EscorciaSalas_MJ_2008,Kocak_PSCB_2010} and have yet to be observed
experimentally.

For the shear modulus $\left(G_\sVRH\right)$, 482 materials (63\% of the test set) are predicted within 25\% 
of calculated values. 
Just one system, C$_{3}$N$_{4}$ (\ICSD\ \#151781), deviates beyond 100 GPa from its calculated value.
The Debye temperature $\left(\theta_\sD\right)$ is predicted to within 50 K accuracy for 540 systems (70\% of the test set). 
BeF$_{2}$ (\ICSD\ \#173557), yet another cage (sodalite) structure~\cite{Zwijnenburg_JACS_2008}, has among the largest errors 
in three models including $\theta_\sD$ (SE = -423 K) and both heat capacities 
{($C_\sp$: SE = 0.65 $k_\sB$/atom; $C_\sV$: SE = 0.61 $k_\sB$/atom)}.
Similar to other ill-predicted structures, this polymorph is theoretical, and has yet to be synthesized.

\begin{figure*}
  \includegraphics[width=\textwidth]{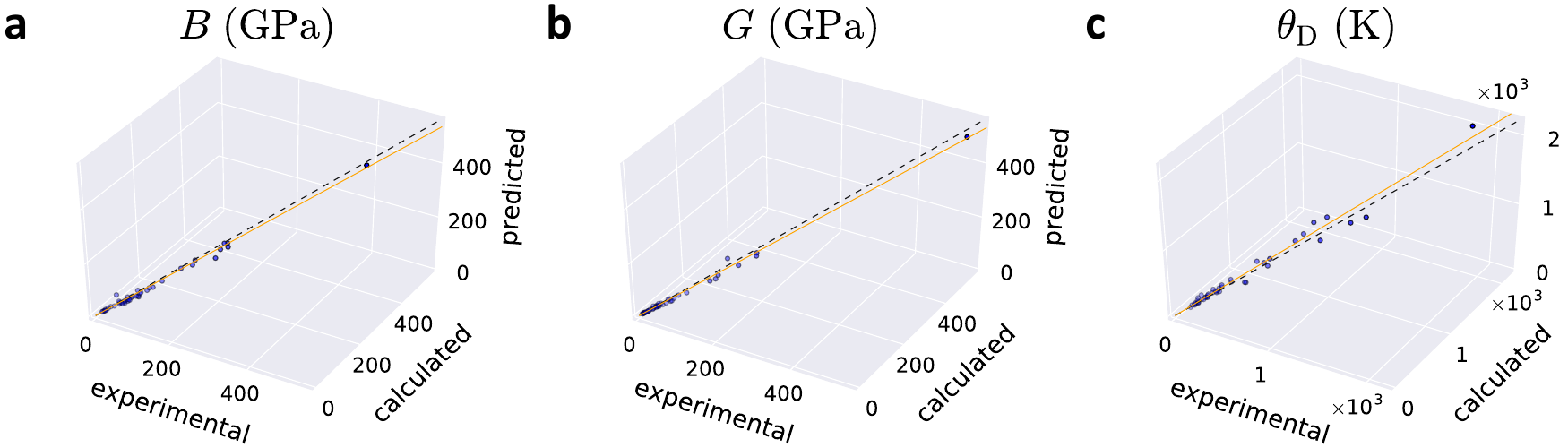}
  \caption{\small 
    \textbf{Comparison of the \AEL-\AGL\ calculations and \ML\ predictions with experimental values for three thermomechanical properties:}
    \textbf{(a)} bulk modulus $\left(B\right)$, 
    \textbf{(b)} shear modulus $\left(G\right)$, 
    and
    \textbf{(c)} Debye temperature $\left(\theta_{\scriptstyle \mathrm{D}}\right)$. 
  }
  \label{figure7}
\end{figure*}

\begin{table*}
\begin{tabular}{|c||c|c||c|c||c|c|}
\hline
\multirow{2}{*}{property} & \multicolumn{2}{c||}{\RMSE} & \multicolumn{2}{c||}{\MAE} & \multicolumn{2}{c|}{$r^{2}$} \\ 
\cline{2-7} 
 & exp. \textit{vs.} calc. & exp. \textit{vs.} pred. & exp. \textit{vs.} calc. & exp. \textit{vs.} pred. & exp. \textit{vs.} calc. & exp. \textit{vs.} pred. \\
\hline \hline
$B$               & 8.90 GPa        & 10.77 GPa   & 6.36 GPa    & 8.12 GPa    & 0.99  & 0.99 \\
\hline
$G$               & 7.29 GPa        & 9.15 GPa    & 4.76 GPa    & 6.09 GPa    & 0.99  & 0.99 \\
\hline
$\theta_\sD$            & 76.13 K        & 65.38 K    & 49.63 K     & 42.92 K     & 0.97  & 0.97 \\
\botrule
\end{tabular}
\caption{\small
    Statistical summary of the \AEL-\AGL\ calculations and 
    \ML\ predictions \textit{vs.} experimental values for three thermomechanical properties. 
    (Figure~\ref{figure7}).
}
\label{table3}
\end{table*}

\noindent \textbf{Comparison with experiments.}
A comparison between calculated, predicted, and experimental results is presented in 
Figure~\ref{figure7}, with relevant statistics summarized in Table~\ref{table3}.
Data is considered for the bulk modulus $B$, shear modulus $G$, and (acoustic) Debye temperature $\theta_\sa$ 
for 45 well-characterized materials with 
diamond (SG\# 227, \AFLOW\ prototype \texttt{A\_cF8\_227\_a}), 
zincblende (SG\# 216, \texttt{AB\_cF8\_216\_c\_a}), 
rocksalt (SG\# 225, \texttt{AB\_cF8\_225\_a\_b}), 
and wurtzite (SG\# 186, \texttt{AB\_hP4\_186\_b\_b}) 
structures~\cite{Morelli_Slack_2006,Semiconductors_BasicData_Springer,Haussuhl_ElasticRocksalt_ZP_1960,Chang_ElasticCaSrBaO_JPCS_1977,Kobiakov_ElasticZnOCdS_SSC_1980,Lam_BulkMod_PRB_1987,curtarolo:art96,Toher_AEL_2016,curtarolo:art121}. 
Experimental $B$ and $G$ are compared to the $B_\sVRH$ and $G_\sVRH$ values predicted here, and
$\theta_\sa$ is converted to the traditional Debye temperature $\theta_\sD=\theta_\sa n^{1/3}$,
where $n$ is the number of atoms in the unit cell. 
All relevant values are listed in the Supplementary Information.

Excellent agreement is found between experimental and calculated values, 
but more importantly, between experimental and predicted results.
With error metrics close to or under expected tolerances from the generalizability analysis, 
the comparison highlights effective experimental confidence in the approach.
The experiments/prediction validation is clearly the ultimate objective of the research presented here.

\section{Discussion}
Traditional trial-and-error approaches have proven ineffective in discovering practical materials.
Computational models developed with \ML\ techniques may provide 
a truly rational approach to materials design.  
Typical high-throughput \DFT\ screenings involve exhaustive 
calculations of all materials in the database, often without 
consideration of previously calculated results.  
Even at high-throughput rates, an average \DFT\ calculation of a medium 
size structure (about 50 atoms per unit cell) takes about 1,170 CPU-hours of 
calculations or about 37 hours on a 32-CPU cores node.
However, in many cases, the desired range of values for the target property is known. 
For instance, 
the optimal band gap energy and thermal conductivity for optoelectronic applications
will depend on the power and voltage conditions of the device~\cite{Figge_APL_2009,Zhou_JACerS_2016}.
Such cases offer an opportunity to leverage previous results and savvy \ML\ models, 
such as those developed in this work, for rapid pre-screening of potential materials.  
Researchers can quickly narrow the list of candidate materials and avoid many extraneous 
\DFT\ calculations---saving money, time, and computational resources.  
This approach takes full advantage of previously calculated results, 
continuously accelerating materials discovery.
With prediction rates of about 0.1 seconds per material, the same 32-CPU cores node can screen
over 28 million material candidates per day with this framework.

Furthermore, interaction diagrams as depicted in Figure~\ref{figure5} offer a pathway to design 
materials that meet certain constraints and requirements.
For example, substantial differences in thermal expansion coefficients among the materials used
in high-power, high-frequency optoelectronic applications leads to bending and cracking of the structure
during the growth process~\cite{Figge_APL_2009,Zhou_JACerS_2016}.
Not only would this work-flow facilitate the search for semiconductors with large band gap energies,
high Debye temperatures (thermal conductivity),
but also materials with similar thermal expansion coefficients. 

While the models themselves demonstrate excellent predictive power with minor deviations, outlier analysis reveals
theoretical structures to be among the worst offenders.
This is not surprising, as the true stability conditions (\textit{e.g.}, high-pressure/high-temperature) have yet
to be determined, if they exist at all.
The \ICSD\ estimates that structures for over 7,000 materials (or roughly 4\%) come
from calculations rather than actual experiment.
Such discoveries exemplify yet another application for \ML\ modeling, rapid/robust curation of large datasets.

To improve large-scale high-throughput computational screening for the identification 
of materials with desired properties, fast and accurate data mining approaches 
should be incorporated into the standard work-flow. 
In this work, we developed a universal \QMSPR\ framework for predicting electronic 
properties of inorganic materials.
Its effectiveness is validated through the prediction of eight key materials properties 
for stoichiometric inorganic crystalline materials, including
the metal/insulator classification, 
band gap energy, bulk and shear moduli, Debye temperature, heat capacity (at constant
pressure and volume), and thermal expansion coefficient.
Its applicability extends to all 230 space groups and the vast majority of 
elements in the periodic table. 
All models are freely available at \href{http://aflow.org/aflow-ml}{aflow.org/aflow-ml}.

\section*{Acknowledgements}
A. T. and O. I. acknowledge support from DOD-ONR (N00014-13-1-0028 and N00014-16-1-2311) and ITS Research Computing Center at UNC.  
Development of the web service was supported by the Russian Scientific Foundation (\# 14-43-00024).  
O. I. acknowledges Extreme Science and Engineering Discovery Environment ({\small XSEDE}) award DMR-110088, 
which is supported by National Science Foundation grant number ACI-1053575.  
S. C. and C. T. acknowledge support from DOD-ONR (N00014-13-1-0030, N00014-13-1-0635), 
DOE (DE-AC02-05CH11231, specifically BES Grant \# EDCBEE), and the Duke University Center for Materials Genomics.  
C.O. acknowledges support from the National Science Foundation Graduate Research Fellowship under Grant No. DGF-1106401.  
\AFLOW\ calculations were performed at the Duke University Center for Materials Genomics.
The authors thank Drs. Mark Asta, Natalio Mingo, Jes\'{u}s Carrete, Kristin
Persson, and Gerbrand Ceder for helpful discussions.

\section*{Author contributions}
A.T. and S.C. designed the study.
O.I. developed and implemented the method.
C.O. and C.T. prepared the data and worked with the \AFLOW\ database. 
E.G. developed the open-access online application available at
\href{http://aflow.org/aflow-ml}{aflow.org/aflow-ml} leveraging the \ML\ models.
O.I and C.O. contributed equally to the work.
All authors discussed the results and their implications and contributed
to the writing of the article.

\newcommand{\Ozolins}{Ozoli\c{n}\v{s}}

\end{document}